\documentstyle[aps,preprint]{revtex}
\begin{document}

\draft

\title{Spectral sum rules for the Tomonaga-Luttinger model}
\author{K. Sch\"onhammer and V. Meden}
\address{
Institut f\"ur Theoretische Physik, Universit\"at G\"ottingen,
Bunsenstrasse 9,\\ D-3400 G\"ottingen, Germany
}
\date{Received 10. May 1993}
\maketitle

\begin{abstract}
\widetext
In connection with recent publications we discuss spectral sum
rules for the Tomonaga-Luttinger model without using the explicit
result for the one-electron Green's function. They are usefull in the
interpretation of recent high resolution photoemission spectra of
quasi-one-dimensional conductors. It is shown that the limit of
infinite frequency and band cut\-off do not commute. Our result
for arbitrary shape of the interaction potential
generalizes an earlier discussion by Suzumura. A general analytical
expression for the spectral function for wave vectors far from the
Fermi wave vector $k_{F}$ is presented.
Numerical spectra are shown to illustrate the
sum rules.
\end{abstract}
\narrowtext

In recent publications the universal \cite{1,2} and the nonuniversal
\cite{3} spectral properties of the Tomonaga-Luttinger (TL) model for
one-dimensional interacting fermions have been studied. The results
were used in the attempt to interpret high resolution photoemission
measurements on quasi-one-dimensional conductors. \cite{4,5,6,6a} For the
discussion of angular integrated spectra a sum rule of the type
proposed by Suzumura \cite{7} is useful. In the following we present
simple derivations of two versions of the sum rule which generalize
Suzumura's result to the case of an interaction potential with arbitrary shape
and do {\it not} require the explicit knowledge of the interacting
Greens function $G^{<}(x,t)$ of the model. Our discussion shows
clearly that the infinite frequency limit and the limit of infinite
momentum cutoff {\it do not commute}. Explicit numerical results
are used to illustrate
this behavior.

We study the TL model with a {\it finite-range} interaction. In the
fermionic representation the kinetic energy $\hat{T}$ and the
interaction term $\hat{V}$ are given by
\begin{eqnarray}
\label{eqn1}
\hat{T} & = &\sum_{k>\Lambda_{+},\sigma} v_{F} k
\hat{a}_{k,+,\sigma}^{\dagger} \hat{a}_{k,+,\sigma}^{} \nonumber \\
&& +
\sum_{k<\Lambda_{-},\sigma}(- v_{F}) k
\hat{a}_{k,-,\sigma}^{\dagger} \hat{a}_{k,-,\sigma}^{} \; ,
\end{eqnarray}
\begin{eqnarray}
\label{eqn2}
\hat{V}=\frac{1}{2} \sum_{\alpha,\alpha',\sigma,\sigma'}
 && \int_{0}^{L} dx \int_{0}^{L} dx' \hat{\psi}_{\alpha,\sigma}^{\dagger}(x)
\hat{\psi}_{\alpha',\sigma'}^{\dagger}(x') \nonumber \\
 && \times v_{\alpha,\alpha'}^{\sigma,\sigma'}(x-x')
\hat{\psi}_{\alpha',\sigma'}^{}(x')
\hat{\psi}_{\alpha,\sigma}^{}(x) \; .
\end{eqnarray}
With finite band cutoffs $\Lambda_{+}$ and $\Lambda_{-}$ the ground state
is well defined as $\hat{H}=\hat{T}+\hat{V}$ has a lower bound. Later
we examine the limits $\Lambda_{+} \rightarrow -\infty$ and
$\Lambda_{-} \rightarrow \infty$. For a zero-range potential,
$v_{\alpha,\alpha'}^{\sigma,\sigma'}(x)=
v_{\alpha,\alpha'}^{\sigma,\sigma'} \delta(x)$
the
interaction terms with $\alpha=\alpha'$ and $\sigma=\sigma'$ (the
$g_{4}^{\scriptscriptstyle{\|}}$-terms in the ``g-ology'' classification
\cite{8})
{\it vanishes} because the product of field operators at the same point
$\left( \hat{\psi}_{\alpha,\sigma}^{}(x) \right)^{2} $
is zero due to the Pauli principle ({\it not} because
$v_{\alpha,\alpha}^{\sigma,\sigma}$ is zero itself). These terms
{\it contribute} for {\it finite-range} interactions. If one wants to describe
interacting nonrelativistic fermions as in Tomonaga's original paper
\cite{9} all interaction terms are equal
$v_{\alpha,\alpha'}^{\sigma,\sigma'}(x-x')=v(x-x')$
where $v$ is the spin independent two body potential.

The spectral function $\rho_{\alpha,\sigma}^{<}(k,\omega)$ relevant for angular
resolved photoemission is given by
\begin{equation}
\label{eqn3}
\rho_{\alpha,\sigma}^{<}(k,\omega)  \equiv
\mbox{$\langle  \phi_{0}^{N}  |  $}\hat{a}_{k,\alpha,\sigma}^{\dagger}
\delta\left(\omega+\left(\hat{H}-E_{0}^{N}\right) \right)
\hat{a}_{k,\alpha,\sigma}^{} \mbox{$| \phi_{0}^{N}  \rangle$}
\end{equation}
and the total spectral weight per unit length is
obtained by a momentum integration
after performing the limit $L \rightarrow \infty$
\begin{equation}
\label{eqn4}
\rho_{+,\sigma}^{<}(\omega) =  \int_{\Lambda_{+}}^{\infty}
\frac{dk}{2\pi} \rho_{+,\sigma}^{<}(k,\omega)  \; .
\end{equation}
The spectral weight is different from zero for $\omega<\mu$, where
$\mu=E_{0}^{N}-E_{0}^{N-1}$ is the chemical potential. For fixed
$\Lambda_{+}$ integration over $\omega$ yields $\int
\rho_{+,\sigma}^{<}(\omega) d \omega=n_{+,\sigma}$, where the density
$n_{+,\sigma}$ is
independent of the interaction strenght. This leads to the (trivial)
first version of the sum rule
\begin{equation}
\label{eqn5}
\lim_{\Lambda_{+} \rightarrow -\infty}
\lim_{\Omega \rightarrow -\infty} \int_{\Omega}^{\infty} d \omega
\left[ \rho_{+,\sigma}^{<}(\omega) -
\left(\rho_{+,\sigma}^{<}(\omega)\right)^{(0)}   \right]=0
\end{equation}
where
$\left(\rho_{+,\sigma}^{<}(\omega)\right)^{(0)}=\Theta(v_{F}
k_{F}-\omega) \Theta(\omega -v_{F} \Lambda_{+})
/(2\pi v_{F})$ is the spectral function for noninteracting
fermions.

A more interesting sum rule can be obtained by commuting the limits
$\Lambda_{+} \rightarrow -\infty$ and $\Omega \rightarrow -\infty$.
For $|\Omega - \mu |$ much larger than $v_{F} k_{c}$, where $1/k_{c}$
is the range of the interaction, one expects $\rho_{+,\sigma}^{<}(\omega)$
to approach $\left(\rho_{+,\sigma}^{<}(\omega)\right)^{(0)}$ such
that the frequency integral in (\ref{eqn5}) reaches a well defined
value independent of $\Omega$ for $|\Lambda_{+}-k_{F}| v_{F} \gg
|\Omega - \mu | \gg v_{F} k_{c}$. This is guaranteed by the fact that
for $k \ll k_{F}-k_{c}$ the {\it shape} of the spectral function
$\rho_{+,\sigma}^{<}(k,\omega)$ becomes {\it independent of the
momentum}, \cite{3} i.e. $\rho_{+,\sigma}^{<}(k,\omega)=f(\omega-v_{F}
k)$, where the function $f$ decays rather quickly away from its
maximum. This property is sufficient to derive the other form of the
sum rule. For a special form of the interaction this shape
independence has been demonstrated in Ref. \onlinecite{3}. We later present
an explicit expression
for the form of the function $f$ for an arbitrary interaction potential.

In order to calculate
\begin{equation}
\label{eqn6}
A_{+,\sigma}(\Omega ) \equiv \int_{\Omega}^{\infty} d \omega
\left[ \rho_{+,\sigma}^{<}(\omega) -
\left(\rho_{+,\sigma}^{<}(\omega)\right)^{(0)}   \right]
\end{equation}
we split the momentum integration into two parts
\begin{eqnarray}
\label{eqn7}
A_{+,\sigma}(\Omega ) &  = &  \int_{\Omega}^{\infty} \left(
\int_{\Lambda_{+}}^{k_{\Omega}} \frac{dk}{2\pi} \delta
\rho_{+,\sigma}^{<}(k,\omega) \right) d \omega  \nonumber \\
&& +
  \int_{\Omega}^{\infty} \left(
\int_{k_{\Omega}}^{\infty} \frac{dk}{2\pi} \delta
\rho_{+,\sigma}^{<}(k,\omega) \right) d \omega
\end{eqnarray}
where  $\delta \rho_{+,\sigma}^{<}(k,\omega) \equiv
\rho_{+,\sigma}^{<}(k,\omega)- \left(
\rho_{+,\sigma}^{<}(k,\omega)\right)^{(0)}$ with
$\left( \rho_{+,\sigma}^{<}(k,\omega)\right)^{(0)}= \Theta(k_{F}-k)
\Theta(k- \Lambda_{+})
\delta(\omega-v_{F} k)$ and $k_{\Omega} \ll k_{F}-k_{c}$ is chosen
such that $\rho_{+,\sigma}^{<}(k,\omega)$ is practically zero for
$\omega < \Omega$. Performing the $\omega$-integration first, yields
for the second term
\begin{equation}
\label{eqn8}
\left( A_{+,\sigma}(\Omega) \right)^{(2)}= \int_{k_{\Omega}}^{\infty}
\frac{dk}{2\pi} \left( n_{k,+,\sigma}^{} - n_{k,+,\sigma}^{(0)} \right) \; .
\end{equation}
As the occupation numbers $n_{k,+,\sigma}$ in the interacting case
approach the noninteracting ones for $k \ll k_{F}-k_{c}$ the lower
integration limit can be replaced by $\Lambda_{+}$, which shows that
$\left( A_{+,\sigma}(\Omega) \right)^{(2)}$ vanishes. In the first
term we can replace $\rho_{+,\sigma}^{<}(k,\omega)$ by $f(\omega -
v_{F} k)$ and change the $k$-integration variable to $x=\omega -v_{F}k$
\begin{equation}
\label{eqn9}
A_{+,\sigma}(\Omega) = \frac{1}{2\pi v_{F}} \int_{\Omega}^{\infty}
\left( \int_{\omega - v_{F} k_{\Omega} }^{\omega - v_{F} \Lambda_{+} }
\left[ f(x)-\delta (x) \right] dx \right) d \omega \; .
\end{equation}
If we perform the limit $\Lambda_{+} \rightarrow -\infty$  the upper
integration limit of the $x$-integration goes to infinity. Now we
partially integrate with respect to $\omega$. The boundary terms
vanish due to the assumptions about $k_{\Omega}$ and the function
$f(x)$ and one obtains
\begin{eqnarray}
\label{eqn10}
A_{+,\sigma}(\Omega) & = & \frac{1}{2\pi v_{F}} \int_{\Omega}^{\infty}
\omega \left[ f(\omega -k_{\Omega} v_{F})  \right. \nonumber \\ && -
\left. \delta ( \omega -k_{\Omega} v_{F})  \right]  d \omega  \\
& = & \frac{1}{2\pi v_{F}} \int_{-\infty}^{\infty} x f(x) dx \; .
\nonumber
\end{eqnarray}
In the second line we have used that the function $f$ is normalized
to unity. This result shows that $A_{+,\sigma}(\Omega)$ is given by
the difference $\delta \mu_{1}^{<}$ of the first moments of the
$k$-dependent spectral functions which is $k$-independent for $k \ll
k_{F}-k_{c}$. We therefore have
\begin{equation}
\label{eqn11}
\lim_{\Omega \rightarrow -\infty}
\lim_{\Lambda_{+} \rightarrow -\infty}  \int_{\Omega}^{\infty} d
\omega \left[ \rho_{+,\sigma}^{<}(\omega) -\left(
\rho_{+,\sigma}^{<}(\omega) \right)^{(0)} \right] =\frac{\delta
\mu_{1}^{<}}{2\pi v_{F}}
\end{equation}
i.e. the limits $\Omega \rightarrow -\infty$ and $\Lambda_{+}
\rightarrow -\infty$ do not commute as mentioned earlier. The
expression on the right-hand side (rhs) of Eq. (\ref{eqn11}) can be calculated
easily,
as for $k \ll k_{F}-k_{c}$ the spectral function
$\rho_{+,\sigma}^{<}(k,\omega)$ equals $\rho_{+,\sigma}^{}(k,\omega)
\equiv \rho_{+,\sigma}^{<}(k,\omega) + \rho_{+,\sigma}^{>}(k,\omega)$
i.e. $\delta \mu_{1}^{<}= \delta \mu_{1}$ where $\delta \mu_{1}$ is
the difference of the first moments of the {\it total} spectral
function $\rho_{+,\sigma}^{}(k,\omega)$, which can easily be
calculated
\begin{eqnarray}
\label{eqn12}
\left( \delta \mu_{1}(k) \right)_{+,\sigma} & = & \left\langle \left\{
\left[ \hat{a}_{k,+,\sigma}^{} \, , \, \hat{V} \right] \, , \,
\hat{a}_{k,+,\sigma}^{\dagger} \right\} \right\rangle  \\
& = &  \sum_{\alpha',\sigma'}
\tilde{v}_{+,\alpha'}^{\sigma,\sigma'}(k=0) n_{\alpha',\sigma'}
\nonumber \\ && -
 \frac{1}{L} \sum_{k'}
\tilde{v}_{+,+}^{\sigma,\sigma}(k-k') n_{k',+,\sigma} \nonumber \\
& \stackrel{k \ll k_{F}-k_{c}}{\longrightarrow} &
  \sum_{\alpha',\sigma'}
\tilde{v}_{+,\alpha'}^{\sigma,\sigma'}(k=0) n_{\alpha',\sigma'}
\nonumber \\ &&-
 \frac{1}{L} \sum_{k'}
\tilde{v}_{+,+}^{\sigma,\sigma}(k') \; . \nonumber
\end{eqnarray}
In order to interpret the sum rule Eq. (\ref{eqn11}) it is necessary
to observe that the chemical potential $\mu$ of the interacting
system and therefore the threshold {\it differs} from the
noninteracting value $\mu^{(0)}=v_{F} k_{F}$. The value of $\mu$ can
be read off the particle number operator dependent terms of
$\hat{V}$. It is given by
\begin{eqnarray}
\label{eqn13}
\mu_{+,\sigma} & = & k_{F}v_{F} +  \sum_{\alpha',\sigma'}
\tilde{v}_{+,\alpha'}^{\sigma,\sigma'}(k=0) n_{\alpha',\sigma'}
\nonumber \\ && -\frac{1}{2L} \sum_{k'} \tilde{v}_{+,+}^{\sigma,\sigma}(k') \;
{}.
\end{eqnarray}
If one artificially shifts the threshold of the unperturbed spectral
density to the {\it same} value as in the interacting case we finally
obtain
\begin{eqnarray}
\label{eqn14}
\lim_{\Omega \rightarrow -\infty}
\lim_{\Lambda_{+} \rightarrow -\infty}  \int_{\Omega}^{\mu} d
\omega \left[ \rho_{+,\sigma}^{<}(\omega) -\frac{1}{2\pi v_{F}} \right]
\nonumber \\ =- \frac{1}{4\pi v_{F}} \int_{-\infty}^{\infty} \frac{dk}{2\pi}
\tilde{g}_{4}^{\scriptscriptstyle{\|}}(k) \; .
\end{eqnarray}
This generalizes the sum rule presented by Suzumura \cite{7}
to the case of an interaction potential with arbitrary shape.
We point out that a straightforward generalization of
Suzumura's approach, which uses an {\it approximate} explicit form of the
Green's function $G_{\alpha,\sigma}(x,t)$, does not in general give
the correct form of the sum rule, as only
$\tilde{g}_{4}^{\scriptscriptstyle{\|}}(k=0)$ and the
value of the interaction cutoff enters. \cite{10} There are various
suggestions in the literature that
$\tilde{g}_{4}^{\scriptscriptstyle{\|}}(k)$ is an odd function of
$k$. This is incorrect, in fact
$\tilde{g}_{4}^{\scriptscriptstyle{\|}}(k)$ is an {\it even}
function of $k$ for the original Tomonaga model and the rhs of Eq.
(\ref{eqn14}) is different from zero if $v(x=0)$ is different from
zero.

How can one understand the difference between the two forms
(\ref{eqn5}) and (\ref{eqn11}) or (\ref{eqn14}) of the sum rule? The
answer is simple and can be illustrated by the numerical results
presented in Ref. \onlinecite{3}: The total spectral weight missing in the low
energy regime is pushed all the way to the {\it lower end of the
spectrum}, which extends beyond the limit for noninteracting electrons.
This part of the weight is accounted for in the form (\ref{eqn5}) but
it is {\it not} if the limits $\Lambda_{+} \rightarrow -\infty$ and
$\Omega \rightarrow -\infty$ are interchanged as in Eq.
(\ref{eqn11}).

We finally illustrate our assumption about the shape independence of
the spectral function $\rho_{+,\sigma}^{<}(k,\omega)$ for momenta $k
\ll k_{F}-k_{c}$ by explicitely presenting an expression for the
function $f$ (which was not necessary to obtain the expression on the
rhs of Eq. (\ref{eqn14}) ).

In the thermodynamic limit the Green's function has the form
\begin{equation}
\label{eqn15}
i G_{+,\sigma}^{<}(x,t)= \frac{-i/2\pi}{x-v_{F} t-i0} e^{i(k_{F}x- \mu
t)}
\exp{\left\{ \tilde{F}(x,t)\right\} }
\end{equation}
where $\tilde{F}(x,t)$ can be calculated e.g. by bosonization.
In the most general case $\tilde{F}(x,t)$ is given by \cite{1}
\begin{eqnarray}
\label{eqn16}
\tilde{F}(x,t) & = & \frac{1}{2}\int_{0}^{\infty} \frac{1}{q} \left[
e^{-iq\left( x-\tilde{v}_{F}^{c}(q)t \right)}-
e^{-iq\left(x-v_{F} t \right) } \right. \nonumber \\
&& \left. +2 s_{c}^{2}(q) \left( \cos{\left(qx\right)}
e^{iq\tilde{v}_{F}^{c}(q)t}-1 \right)  \right] dq  \nonumber \\
&& + \frac{1}{2} \int_{0}^{\infty} \frac{1}{q} \left[c \rightarrow s \right]
dq
\end{eqnarray}
where $\tilde{v}_{F}^{c/s}(q)=\left[ \left(
v_{F}+\tilde{g}_{4}^{c/s}(q)/\pi\right)^{2}-
\left( \tilde{g}_{2}^{c/s}(q)/\pi\right)^{2} \right]^{1/2}$ is the Fermi
velocity
of the charge (spin) degrees of freedom, $1+2s_{c/s}^{2}(q)=\left( v_{F}
+\tilde{g}_{4}^{c/s}(q)/\pi \right) /\tilde{v}_{F}^{c/s}(q)$ and
$\tilde{g}_{i}^{c/s}(q)=\left(
\tilde{g}_{i}^{\scriptscriptstyle{\|}}(q) \pm \tilde{g}_{i}^{\perp}(q)
\right)/2$.
Under rather weak
assumptions on the $\tilde{g}_{i}^{\nu}(k)$ ($\nu=
\displaystyle{\perp}$,
$\|$)
the function $\exp{\left[
\tilde{F}(x,t) \right] }$ is analytic in $x$ in a strip around the
real axis. If the Fourier transform of $G_{+,\sigma}^{<}(x,t)$ with
respect to the $x$-variable is performed by contour integration, only
the pole at $v_{F}t+i0$ contributes for $k \ll k_{F}-k_{c}$. This
yields the shape independence and the explicit result for the shape
function in the form of a Fourier integral
\begin{equation}
\label{eqn17}
\rho_{+,\sigma}^{<}(k,\omega) = \frac{1}{2\pi} \int_{-\infty}^{\infty}
dt e^{i(\omega - v_{F} k)t} e^{i(v_{F} k_{F} -\mu)t}
e^{\tilde{F}(v_{F}t,t)} \; .
\end{equation}
In contrast to the method presented in Ref. \onlinecite{3} this
result allows to calculate the shape function $f(\omega)$ for {\it
arbitrary} potentials. Singularities of the spectrum or its
derivatives can be obtained by the large $t$ asymptotic expansion of
the momentum integral in (\ref{eqn16}), which determines the function
$\tilde{F}(v_{F} t,t)$. For smooth potentials the frequencies
$\omega^{c/s}(q_{0}^{c/s})=q_{0}^{c/s}
\tilde{v}_{F}^{c/s}(q_{0}^{c/s})$ with $q_{0}$ determined by
the stationary phase condition $\left( \partial \omega / \partial q
\right) |_{q_{0}}=v_{F}$ lead to prominent features in the spectral
function.
A numerical
calculation of the shape function $f$  for a spin independent two
particle interaction
($\tilde{g}_{4}^{\scriptscriptstyle{\|}}(q)
\equiv \tilde{g}_{4}^{\perp}(q) \equiv \tilde{g}_{4}(q) $,
$\tilde{g}_{2}^{\scriptscriptstyle{\|}}(q)
\equiv \tilde{g}_{2}^{\perp}(q) \equiv \tilde{g}_{2}(q) $)
is shown in Fig. 1 for
repulsive interactions
$\tilde{v}(q)=\tilde{g}_{4}(q)=\tilde{g}_{2}(q) \geq 0$. The full
curve corresponds to a step, the dotted line to a Gaussian
function for $\tilde{v}(q)$ and the dashed curve to an approximation
discussed below. As $\tilde{v}_{F}(q)\geq v_{F}$ for all
values of $q$ the spectra have a sharp threshold.
For the Gaussian potential the first derivative is singular at
$\omega-v_{F}k=-\omega^{c}(q_{0}^{c})$ and has a discontinuity at
$-2\omega^{c}(q_{0}^{c})$. For the step potential the first
derivative shows a discontinuity at $\omega-v_{F} k=-v_{F} k_{c}$ as
obtained by a different method in Ref. \onlinecite{3}. These spectral
functions are {\it exact} within accuracy of the drawings.
{\it Approximate} results for the Green's function $G_{+,\sigma}^{<}(x,t)$
and corresponding results for spectral functions have been obtained
by various authors \cite{11,7,10} by replacing the momentum integral
in (\ref{eqn16}) for an interaction potential of range $r=1/k_{c}$ by
an exponential cutoff
$\int_{0}^{\infty}  \exp{(-q/k_{c})} \ldots dq$ and the $q$-dependent
interactions by their $q=0$ values. By this approximation essential
features of the spectra are lost as the dashed curve in Fig. 1 shows.
In Fig. 2 we present the
corresponding results for the total spectral weight
$\rho_{+,\sigma}^{<}(\omega)$, which are {\it monotonic} functions of
frequency. This is in contrast to the model \cite{7} of a spin
dependent interaction with $\tilde{g}_{4}^{\scriptscriptstyle{\|}}(q) \equiv
0$,
respectively a pure interbranch interaction
(spin dependent or spin independent)
with $\tilde{g}_{4}^{\scriptscriptstyle{\|}}(q) \equiv \tilde{g}_{4}^{\perp}(q)
\equiv
0$, which are unphysical assumptions for finite-range interactions. \cite{3}

\begin{figure}
\caption{Shape function $f$ for the step interaction (solid line),
the Gaussian interaction (dotted line) and the approximation which is
independent of the details of the interaction (dashed line)
for $\tilde{v}_{F}(q=0)=2
v_{F}$. The energy is measured relative to $\mu -v_{F} k_{F}$.}
\vspace{2.0cm}
\caption{Total spectral weight $\rho_{+,\sigma}^{<}$ for the step
interaction (solid line) and the Gaussian interaction (dotted line)
for $\tilde{v}_{F}(q=0)=3 v_{F}$. The dashed curve shows again the
discussed approximation. The energy is measured relative to
$\mu$.}
\end{figure}
\end{document}